\documentclass
[preprint,showpacs,byrevtex,10pt,twocolumn,tightenlines,prl,letterpaper]{revtex4}%
\usepackage{amsfonts}
\usepackage{amsmath}
\usepackage{amssymb}
\usepackage[dvips]{graphicx}
\usepackage{hyphenat}%
\providecommand{\U}[1]{\protect\rule{.1in}{.1in}}

\begin{document}
\preprint{ }
\title{Coexistence of Electron-Glass Phase and Persistent Photoconductivity in GeSbTe Compounds}
\author{Z. Ovadyahu}
\affiliation{Racah Institute of Physics, The Hebrew University, Jerusalem 91904, Israel }


\begin{abstract}
It is demonstrated that persistent-photoconductivity (PPC), well-studied in
lightly-doped semiconductors, is observable in GeSbTe compounds using infrared
excitation at cryogenic temperatures. The low level of energy-flux necessary
to induce an appreciable effect seems surprising given the high
carrier-concentration n of these ternary alloys (n%
$>$%
10$^{\text{20}}$cm$^{\text{-3}}$). On the other hand, their high density of
carriers makes GeSbTe films favorable candidates for exhibiting intrinsic
electron-glass effects with long relaxation times. These are indeed observed
in GeSbTe thin-films that are Anderson-localized. In particular, a memory-dip
is observed in samples with sheet resistances larger than $\approx
$10$^{\text{5}}\Omega$ at T$\approx$4K with similar characteristics as in
other systems that exhibit intrinsic electron-glass effects.
Persistent-photoconductivity however is observable in GeSbTe films even for
sheet resistances of the order of 10$^{\text{3}}\Omega$, well below the range
of disorder required for observing electron-glass effects. These two
non-equilibrium phenomena, PPC and electron-glass, are shown to be of
different nature in terms of other aspects as well. In particular, their
relaxation dynamics is qualitatively different; the excess conductance
$\Delta$G associated with PPC decays with time as a stretched exponential
whereas a logarithmic relaxation law characterizes $\Delta$G(t) of all
electron-glasses studied to date. Surprisingly, the magnitude of the
memory-dip is enhanced when the system is in the PPC state. This
counter-intuitive result may be related to the compositional disorder in these
materials extending over mesoscopic scales. Evidence in support of this
scenario is presented and discussed.

\end{abstract}
\maketitle

\section{Introduction}

Many phenomena in solid state systems exhibit monotonous conductance changes,
suggestive of a non-equilibrium phenomenon. Annealing of defects during
heat-treatment, irradiation by an external source, diffusion of injected
particles, and illumination by light, are familiar examples of such phenomena.

There are two specific phenomena that exhibit sluggish conductance relaxation
following excitation by a light-source: persistent-photoconductivity (PPC)
that has been studied extensively as of 1968
\cite{1,2,3,4,5,6,7,8,9,10,11,12,13}, and the less familiar (but just as
controversial) electron-glass \cite{14,15,16,17,18,19,20,21,22,23}. In
addition to the similar way these effects are reflected in the sample
conductance once triggered by light, both become more prominent at lower
temperatures and both exhibit a non-exponential relaxation. This sometimes led
to assuming a common mechanism to these phenomena \cite{24}.

In this work we describe some light-induced non-equilibrium transport
properties in thin films of GeSb$_{\text{x}}$Te$_{\text{y}}$. The study was
initiated as a test of the conjecture that intrinsic electron-glass effects,
with long relaxation-times, are generic and should be observable in all
Anderson insulating systems with sufficiently large carrier-concentrations
\cite{25,26}. The ternary compound GeSb$_{\text{x}}$Te$_{\text{y}}$ reported
by several groups to have carrier-concentration n in the range $\approx
$10$^{\text{20}}$-10$^{\text{21}}$cm$^{\text{-3}}$ \cite{27,28,29} appeared to
be a suitable candidate for this test. This expectation turned out to be
correct; in their insulating state, GeSb$_{\text{x}}$Te$_{\text{y}}$ films do
exhibit electron-glass behavior. However, a surprisingly large
photosensitivity was encountered in the very first experimental run. The
sample was initially tested by a short exposure to infrared radiation using a
power level that in other electron-glasses yielded a few percent increase in
conductance G. In the GeSb$_{\text{x}}$Te$_{\text{y}}$ sample this resulted in
the conductance-reading promptly going off-scale. It soon became clear that
this photo-response is a manifestation of persistent-photoconductivity rather
than the anticipated electron-glass phenomenon. Persistent-photoconductivity
is referred to in this work for processes that involve charge generation by
exposure to light without change of stoichiometry; this should not be confused
with instances where self-doping is affected when e.g., oxygen is expelled
from the sample by exposure to high-energy photons as happens in
In$_{\text{2}}$O$_{\text{3-x}}$ \cite{30} and ZnO \cite{31}. The latter
processes may appear to be similar in that they show sluggish conductance
relaxations, but they are of a different nature.

Persistent-photoconductivity typically occurs in systems with
carrier-concentration smaller than $\approx$10$^{\text{19}}$cm$^{\text{-3}}$
like lightly-doped semiconductors \cite{1,2,3,4,5,6,7,8,9,10,11,12,13}.
Intrinsic electron-glass effects were never observed in these low-density
systems \cite{26}.

Effects associated with the electron-glass phase were duly observed in
field-effect experiments on Anderson-localized GeSb$_{\text{x}}$Te$_{\text{y}%
}$ samples. These revealed a memory-dip with the same characteristic features
found in previously studied electron-glasses. GeSb$_{\text{x}}$Te$_{\text{y}}$
is the first narrow-gap semiconductor that exhibits intrinsic electron-glass
behavior. By \textquotedblleft intrinsic" we mean that the glassy effects
appear in a given substance \textit{independently} of the way the sample was
prepared to achieve the required parameters (resistance at the measuring
temperature, carrier-concentration, and dimensionality). Most importantly, the
system has to exhibit a memory-dip with a width that is commensurate with the
carrier-concentration of the material \cite{26} .

That electron-glass phase and persistent-photoconductivity coexist in the
GeSb$_{\text{x}}$Te$_{\text{y}}$ compounds gives us a unique tool to learn
about both phenomena. The current study demonstrates that PPC and
electron-glass effects are different phenomena. Using the empirically known
behavior of the electron-glass and persistent-photoconductivity it is argued
that some of the non-trivial transport effects observed when the two coexist
may be a result of the inhomogeneous nature of the system.

\section{Experimental}

\subsection{Sample preparation and characterization}

Samples used for this study were prepared by e-gun depositing a
GeSb$_{\text{2}}$Te$_{\text{5}}$ alloy unto room temperature substrates in a
high-vacuum system (base pressure 0.8-1$\cdot$10$^{\text{-7}}$mbar) with rates
of 2-3\AA /second. A constant thickness of 120\AA ~was used for all films in
this study. Two types of substrates were used; 1mm-thick microscope
glass-slides, and 0.5$\mu$m SiO$_{\text{2}}$ layer thermally grown on
$<$%
100%
$>$
silicon wafers. The Si wafers were boron-doped with bulk resistivity
$\rho\simeq$ 2$\cdot$10$^{\text{-3}}\Omega$cm, deep into the degenerate
regime. These wafers were employed as the gate electrode in\ the field-effect
measurements. The field-effect technique was used heavily in this study as an
analytical tool. The microscope glass-slides were mostly used for optical
characterization and for Hall-Effect measurements, both performed at room-temperatures.

Each deposition batch included samples for optical excitation measurements,
samples for Hall-effect measurements, and samples for structural and chemical
analysis. For the latter study, carbon-coated Cu grids were put close to the
sample during its deposition. The TEM grids received the same post-treatment
as the samples used for transport measurements.

The source material was 99.999\% pure GeSb$_{\text{2}}$Te$_{\text{5}}$
(American Elements). Different preparation runs however, produced films that
usually had different stoichiometry than the source material. In the samples
used in this work the GeTe atomic ratio, for example, varied between 1:3.5 to
1:4.8. These films should therefore be referred to as GeSb$_{\text{x}}%
$Te$_{\text{y}}$. By itself, variations in film composition seem to make
little difference to the transport properties; we were not able to identify
a\ qualitative influence of the film global stoichiometry on the transport
results reported below, the main effect on the latter turns out to be the film
sheet-resistance R$_{\square}$.

Transmission-electron-microscopy (TEM), using the Philips Tecnai F20 G2) were
employed to characterize the films composition and microstructure. The Varian
Cary-1 spectrophotometer was used for optical measurements.

Polycrystalline samples of GeSb$_{\text{x}}$Te$_{\text{y}}$ were obtained by
subjecting the as-deposited (amorphous) films to heat-treatment at$\ $%
a~temperature T$_{\text{H}}$ for 2-3 minutes during which the sample was
crystallized. The difference in the optical properties in the visible due to
the amorphous-crystalline transformation could be seen in films deposited on
glass substrates as a mild change in the color tint. An example is shown in
Fig.1.%
\begin{figure}[ptb]%
\centering
\includegraphics[
height=2.4543in,
width=3.339in
]%
{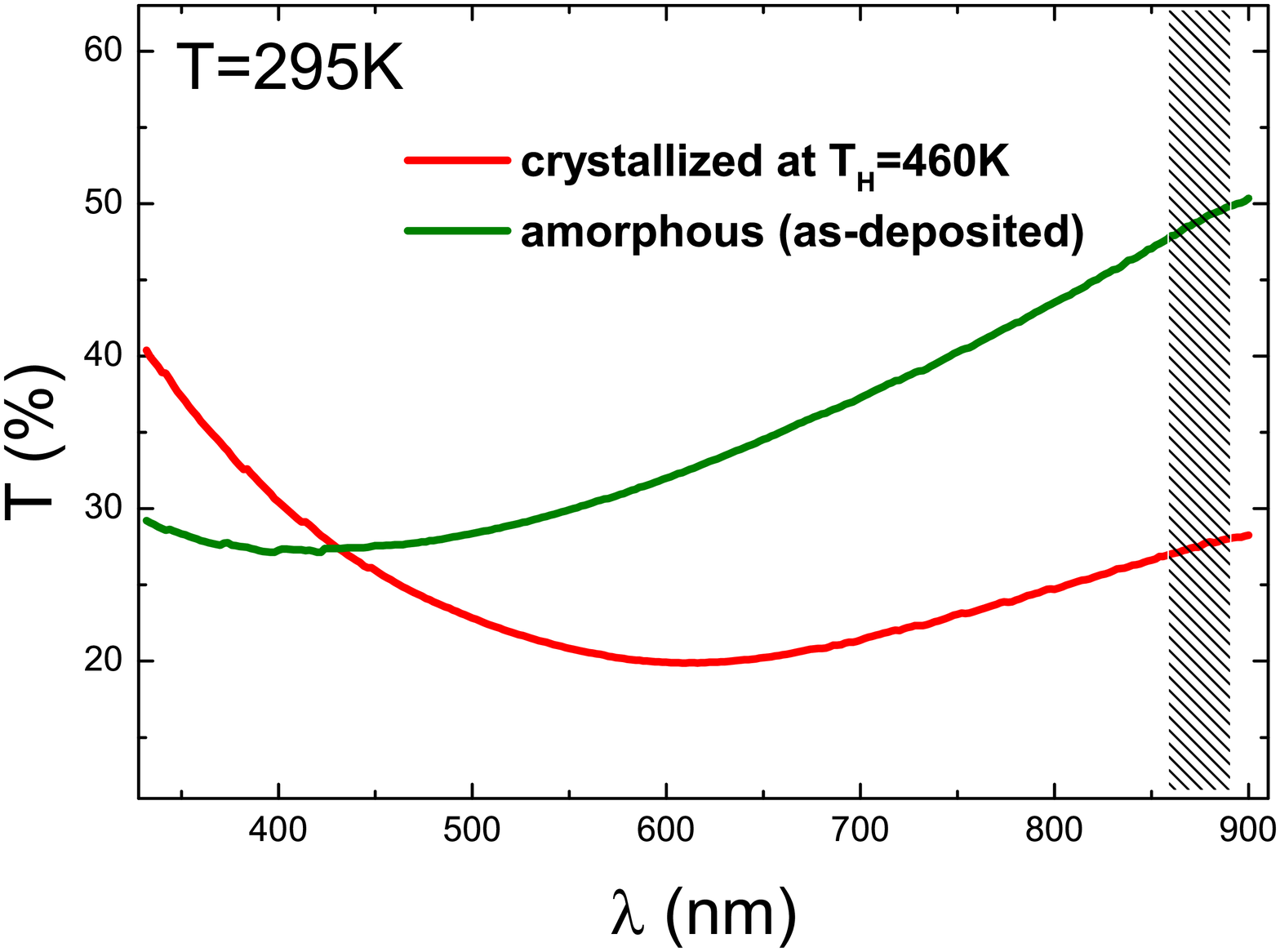}%
\caption{The optical transmission in the visible (relative to the bare
glass-slide) of the as-prepared amorphous and the crystallized film (at
T$_{\text{H}}$=455K). The hatched area indicates the spectral range of the IR
LED used in this work. The absorption of the GeSb$_{\text{x}}$Te$_{\text{y}}$
film at these wavelengths is smaller than 10\%. The difference in the
transmission between the two films is due to stronger reflection, absorption,
and scattering of the crystalline material. }%
\end{figure}

The heat-treatment temperature was held constant during the heat-treatment
process to $\pm$2K around a specific value T$_{\text{H}}$. This was limited to
the range 430-490K for the samples used in this work. In agreement with
reports by other groups \cite{28,32}, the resistivity of a given crystalline
film was found to depend on the value of T$_{\text{H}}$. Using a heating-stage
attachment in the TEM revealed that onset of crystallization could be as low
as 400-410K, and that the grain size increases monotonically with
T$_{\text{H}}$.

A TEM micrograph and associated diffraction pattern of a typical film used in
this study is given in Fig.2. This illustrates the polycrystalline nature of
the film and a tight, space-filling packing of the crystallites with rare
occasional cracks between adjacent crystallites. Grain-sizes as large as
$\approx$10nm start appearing at T$_{\text{H}}\approx$440K, and grain-sizes as
big $\approx$1 micron were observed for T$_{\text{H}}\approx$540K.
Interestingly, at this grain-size, which is much larger than the film
thickness (d=120\AA ), the film still maintains its mosaic, physically
continuous structure

The diffraction pattern taken from these films were consistent with the
rhombohedral (R-3m) phase of GeTe in all our samples. No diffraction ring was
found that could not be related to this basic structure. Note however that the
presence of some amorphous component is always a possibility which is hard to
rule out in these cases (especially when the film is supported on amorphous
carbon). A small amount of texture (preferred orientation) was observed in
these films (see diffraction pattern in Fig.2). There are quite a number of
defects that are observed in the micrograph; grain boundaries and twinning
being the most prevalent. These defects, as well as the
compositional-disorder, discussed later, and surface scattering are presumably
responsible for the low mobility of the films (a desirable feature in the
context of this work).
\begin{figure}[ptb]%
\centering
\includegraphics[
height=3.3399in,
width=3.3399in
]%
{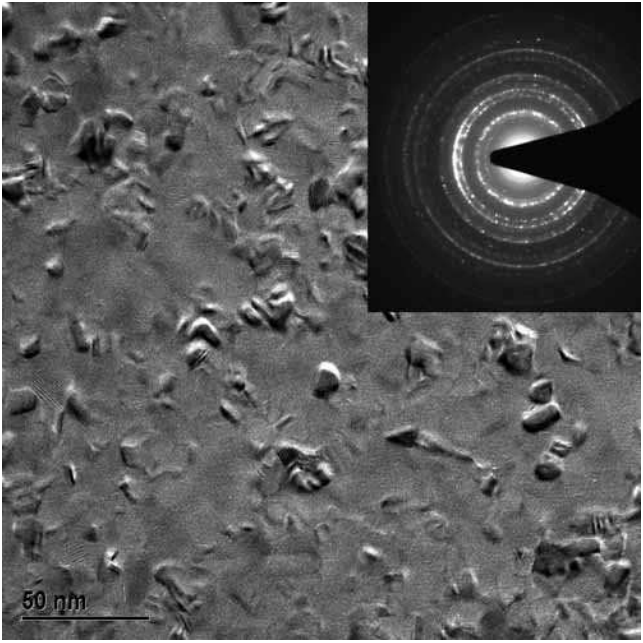}%
\caption{Bright-field image and the associated diffraction pattern of a
typical GeSb$_{\text{x}}$Te$_{\text{y}}$ film crystallized at T$_{\text{H}}%
$=455K. The diffraction in this case was taken at 10~degrees inclination to
expose the film texture.}%
\end{figure}

Gold strips were deposited on the ion-bombarded glass substrates prior to the
GeSb$_{\text{x}}$Te$_{\text{y}}$ deposition. Lateral dimensions of the samples
used here were 0.5-1mm long and 1mm wide. Samples had sheet resistance
R$_{\square}$ in the range 2k$\Omega$-25M$\Omega$ at 4K. This range includes
diffusive samples (R$_{\square}$%
$<$%
h/e$^{\text{2}}$) as well as samples well into the hopping regime
(R$_{\square}{}\gg$h/e$^{\text{2}}$) at the measurement temperature. The
room-temperature carrier-concentration n of these films varied in the range of
(4-9)$\cdot$10$^{\text{20}}$cm$^{\text{-3}}$, a range too narrow to show a
definite correlation with the sample resistance.

\subsection{Measurement techniques}

Conductivity of the samples was measured using a two terminal ac technique
employing a 1211-ITHACO current preamplifier and a PAR-124A lock-in amplifier.
All measurements were performed with the samples immersed in liquid helium at
T$\approx$4.1K held by a 100 liters storage-dewar. This allowed up to two
months measurements on a given sample while keeping it cold (and in the dark)
which was utilized to extend the time-duration of relaxation processes.

The ac voltage bias in conductivity measurements was small enough to ensure
near-ohmic conditions (except for the current-voltage plots). Optical
excitations in this work were accomplished by exposing the sample to an AlGaAs
diode operating at $\approx$0.88$\pm$0.05$\mu$m (see Fig.1), mounted on the
sample-stage typically $\approx$10-15mm from the sample. The diode was
energized by a computer-controlled Keithley 220 current-source.

\section{Results and discussion}

\subsection{Non-equilibrium induced by optical-excitation in GeSb$_{\text{x}}%
$Te$_{\text{y}}$}

The experimental protocol used for optical excitation is illustrated in Fig.3
using a 120\AA \ film of GeSb$_{\text{x}}$Te$_{\text{y}},$ deep in the
insulating regime. The experiment begins $\approx$24 hours after the sample is
cooled-down to 4.1K by recording G(t) for 1-2 minutes to establish a baseline,
near-equilibrium conductance G$_{0}$. The IR source was then turned on for 3
seconds, then turned off while G(t) continues to be measured. The brief IR
burst causes G to promptly increase by $\delta$G$_{\text{IR}}$ which decays
slowly with time once the source is turned off (Fig.3).%
\begin{figure}[ptb]%
\centering
\includegraphics[
height=2.6697in,
width=3.3399in
]%
{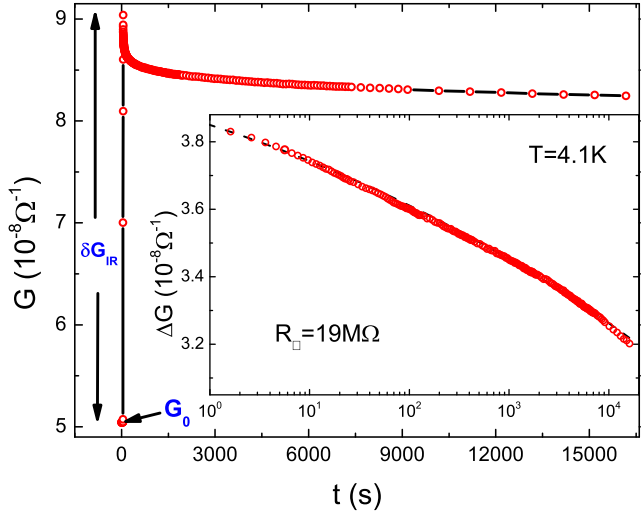}%
\caption{The IR-protocol used on a 19M$\Omega$ sample to illustrate the
response of the conductance and its relaxation. The IR source is turned on for
3 seconds at an intensity of 1mWatt then turned off. The inset shows the
relaxation of the excess conductance as function of time illustrating that the
experimental curve systematically deviates from the logarithmic dependence
(the origin of the time axis in this plot is taken as 1sec after the IR source
is turned off). The dashed line is a mathematical fit to $\delta$G(t)$\propto
$exp[-(t/$\tau$)$^{\beta}$] with $\beta$=0.1 and $\tau$=4$\cdot$10$^{\text{9}%
}$sec.}%
\end{figure}

This change in the sample conductance has a rather small effect on the
conductance versus voltage characteristics as shown in Fig.4. The deviations
from Ohm's law at 4.1K start to become substantial at a field of $\approx
$500V/m, which is similar to what is observed in other disordered films of
comparable resistance at these temperatures \cite{26}. However, these
deviations remain rather small at higher fields possibly suggesting a
relatively small hopping length.
\begin{figure}[ptb]%
\centering
\includegraphics[
height=2.6662in,
width=3.3399in
]%
{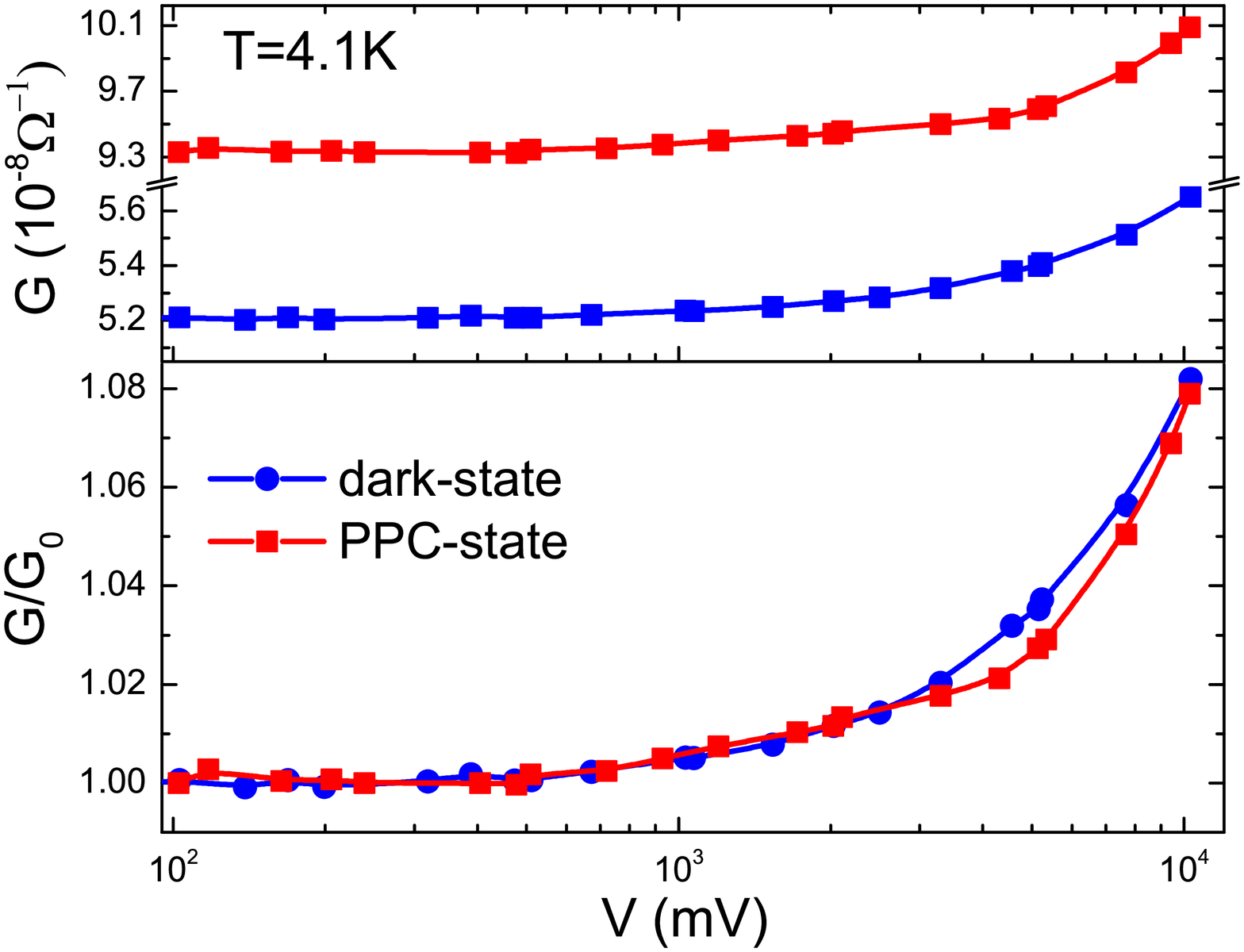}%
\caption{The dependence of the conductance of a GeSb$_{\text{x}}$%
Te$_{\text{y}}$ film (the same sample as in Fig.3) on the applied voltage at
the dark-state and several hours after IR exposure. }%
\end{figure}

On first sight, the behavior of G(t) depicted in Fig.3 appears to be similar
to that of the "IR-protocol" applied to an electron-glass sample (see, for
example, Fig.1 of \cite{33}). A closer look reveals some differences. First,
the effect is much larger than what is usually encountered with any\ of the
previously studied electron-glass; note that a $\approx$3mJ of infrared flux
was sufficient to obtain $\delta$G$_{\text{IR}}$/G$_{\text{0}}\approx$80\%
(Fig.3). For comparison, an energy-flux of $\approx$80mJ achieved only
$\delta$G$_{\text{IR}}$/G$_{\text{0}}$$\simeq$15\% in the most IR-responsive
electron-glass studied to date (see, Fig.6 of \cite{33}).

Secondly, the initial fast drop of the excess conductance (once the IR source
is off), common to all IR-excitations involving electron-glasses \cite{33}, is
not observed here. More importantly, the temporal relaxation-law of G(t) is
different: The inset to Fig.3 demonstrates that the slow component of the
induced excess conductance relaxes with time as a stretched exponent,
G(t)$\propto$exp[-(t/$\tau$)$^{\beta}$]. This should be contrasted with the
logarithmic law for G(t) that is a characteristic feature of the intrinsic
electron-glass \cite{26,34}. The stretched exponential relaxation law has been
consistently observed in more than twenty GeSb$_{\text{x}}$Te$_{\text{y}}$
samples measured in this study. These included samples with sheet resistance
R$_{\square}$ covering the range 2.5k$\Omega$ to 24M$\Omega$ at T$\approx$4K
all showing this relaxation law with the \textit{same} $\beta$ =0.1$\pm$0.005
as a best-fit parameter. Similar values for $\beta$ are typically found at
these temperatures in the relaxation law of the PPC in AlGa$_{\text{x}}%
$As$_{\text{1-x}}$ compounds \cite{5}. The other fit parameter in the
stretched-exponential $\tau$, turned out to be in the range of 10$^{\text{8}%
}\sec$ to 10$^{\text{10}}\sec$,~and showed no systematic correlation with
R$_{\square}$. Apparently the disorder (assessed by the value of the sample
R$_{\square}$) affects this phenomenon mainly through the relative magnitude
of the excess conductance.

Whether the stretched exponential law is an inherent property of this
phenomenon or is just a concise way to parametrize the data is open to debate
(it may be possible to fit $\delta$G(t) by 2-3 exponential terms \cite{11}).
What is clear however is that the functional form of the relaxation deviates
appreciably from the well-established logarithmic relaxation law of intrinsic
electron-glasses \cite{26}. This is obvious in the extended measurements that
were taken here, but might be missed in runs that are shorter than two-decades
in time (due to the rather small value of $\beta$).

It is noteworthy that this nonequilibrium effect persists throughout a wide
range of disorder, including both the insulating and the diffusive regime.
This is illustrated in Fig.5 for a series of GeSb$_{\text{x}}$Te$_{\text{y}}$
films with different degrees of disorder but measured under identical
IR-protocol conditions. This observation, by itself, is a strong evidence that
the $\delta$G$_{\text{IR}}$ is not an electron-glass effect; the latter is
strictly a hopping-regime phenomenon and vanishes as the diffusive regime is
approached \cite{26,35,36}.%
\begin{figure}[ptb]%
\centering
\includegraphics[
height=2.6662in,
width=3.3399in
]%
{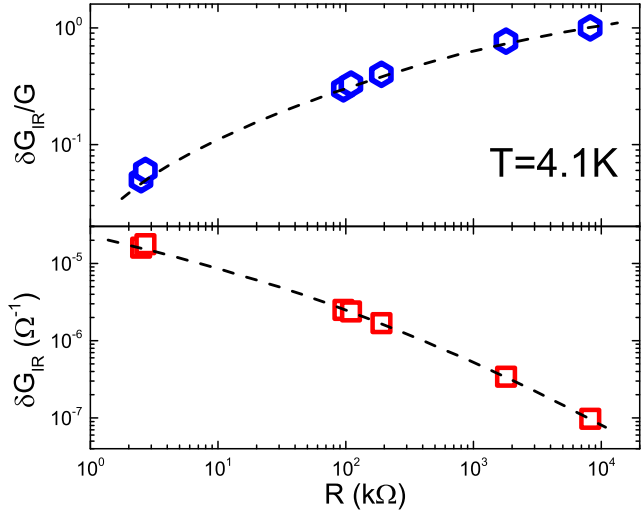}%
\caption{The dependence of the PPC magnitude on film disorder (represented by
R$_{\square}$). These are based just on samples where care was taken to use
identical illumination conditions. See Fig.3 for the definition for $\delta
$G$_{\text{IR}}$ and \ G$_{\text{0}}.$}%
\end{figure}

Actually, the effects described in Fig.3 and Fig.5 are manifestations of a
well-known phenomenon - persistent-photoconductivity (PPC). This phenomenon
has been studied for more than four decades, mostly in lightly-doped
semiconductors. In the majority of cases, the relaxation law was found to be
stretched-exponential \cite{2,4,5,6,7,8,9,10,11,12,13} as in this work.
Different models were offered to account for PPC but\ they all share a common
ingredient; a barrier that hinders the recombination of the photo-created
electron-hole pairs. In essence, the excess conductance due to this mechanism
is an enhanced \textit{carrier-concentration }(in the hopping system $\delta
$G$_{\text{IR}}$may depend exponentially on $\delta$n). The excess conductance
in the electron-glass on the other hand is essentially due to enhanced
\textit{mobility}; the G(t) relaxation reflecting dressing of particles to
form quasi-particles \cite{37}.

It is rather surprising to find PPC in degenerate semiconductors such as these
GeSb$_{\text{x}}$Te$_{\text{y}}$ compounds with their high
carrier-concentration. To the best of our knowledge, this is the first
occurrence of PPC in a system with carrier-concentration n%
$>$%
10$^{\text{20}}$cm$^{\text{-3}}$. Typical values for n in systems exhibiting
PPC range from10$^{\text{15}}$cm$^{\text{-3}}$to 10$^{\text{19}}%
$cm$^{\text{-3}}$\cite{3,7}{\large .} These relatively low
carrier-concentrations made it feasible to demonstrate, via Hall effect
measurements, that the effect is associated with increase of
carrier-concentration \cite{3,7}. High carrier-concentration, on the other
hand, makes the GeSb$_{\text{x}}$Te$_{\text{y}}$ a prime candidate to exhibit
intrinsic electron-glass effects, which was the original reason for choosing
this system for the study.

In the following we present further experimental results demonstrating that
electron-glass effects and PPC are distinct phenomena. They may be separated
by controlling the conditions required for the appearance of each in turn.
They may also coexist, making it possible to learn something on the material
from their interplay.

\subsection{GeSb$_{\text{x}}$Te$_{\text{y}}$ is an intrinsic Electron-Glass
(when Anderson localized)}

It is straightforward to eliminate the PPC - just not turn it on! This should
leave the scene to the electron-glass, which may be probed while in the
"dark-state" by non-optical means. A dark-state of a pre-exposed sample may be
restored by raising it above the helium bath and holding it for few minutes at
T$\eqslantgtr$30K while in the shielded Dewar. T $\approx$30K is presumably of
the order of the energy-barrier associated with the PPC in this system.
Thermal recycling (from 4K to 30-40K) was used to reinstate the sample in its
dark-state if it was in the PPC state. This restored the original conductance
and the field-effect characteristics the sample had prior to IR-exposure to a
very high accuracy, which suggests that there are no irreversible material
changes in the sample due to the IR illumination.

The acid test for intrinsic electron-glass is the existence of a memory-dip
\cite{26}. This involves a field-effect measurement; recording the conductance
as function of a gate-voltage V$_{\text{g}}$. Results of such measurements are
shown below (Fig.6) for the same sample that was used in Fig.3.%
\begin{figure}[ptb]%
\centering
\includegraphics[
height=2.4068in,
width=3.339in
]%
{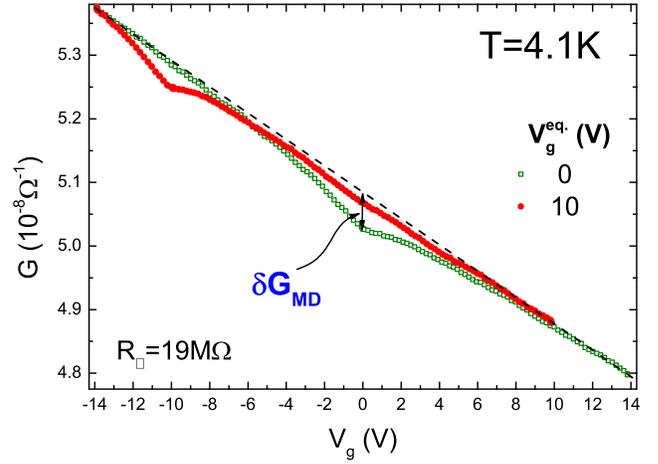}%
\caption{Conductance as function of the gate voltage for the same sample as in
Fig.3 (but here in its dark-state). The history of this protocol described in
the text. The resulting field effect curves G(V$_{\text{g}}$) demonstrate the
characteristic feature of intrinsic electron-glass; a memory dip that appears
as a local minimum around the gate-voltage V$_{\text{g}}^{\text{eq}}$ under
which the sample was allowed to equilibrate. The dashed line is the
thermodynamic component of the field-effect.}%
\end{figure}

Two traces are shown in Fig.6; the first was taken 20 hours after the sample
was cooled down and allowed to equilibrate while V$_{\text{g}}$=0V was held
between the sample and the gate. The resulting field-effect trace
G(V$_{\text{g}}$) is composed of two components; an anti-symmetric component
reflecting the underlying (thermodynamic) density-of-states (DOS), and a
superimposed dip, centered around the V$_{\text{g}}$ where the systems was
allowed to relax (V$_{\text{g}}$=0V in this particular case). The latter, the
so called memory-dip, is the distinguishing feature of the intrinsic
electron-glass \cite{26}. The width of the memory-dip is comparable with that
of the amorphous version of indium-oxide with a comparable
carrier-concentration \cite{38}. The second trace shown in Fig.6 was taken
after the gate-voltage was moved to V$_{\text{g}}$=-10V and left there
overnight. A new G(V$_{\text{g}}$) trace was taken, 21 hours after letting the
sample equilibrate under V$_{\text{g}}$=-10V, to produce the curve labeled by
full circles. This field-effect shows the same thermodynamic component as the
old one while the memory-dip is now centered around the newly imposed
gate-voltage. This illustrates the "two-dip-experiment" another common feature
of electron-glasses \cite{26}.

Note that $\partial$G/$\partial$V$_{\text{g}}$ of the thermodynamic component
has the opposite sign to that observed in n-type semiconductors such as
indium-oxide and thallium-oxide films \cite{26}. This is consistent with the
p-type nature of GeSb$_{\text{x}}$Te$_{\text{y}}$, where the Fermi energy lies
at the valence-band \cite{39}.

A characteristic feature observed in all intrinsic electron-glasses is a
monotonous dependence of the memory-dip (MD) magnitude on disorder (at fixed
sweep-rate and temperature) and, in particular, that it vanishes at the
diffusive regime \cite{26,35,36}. The data in Fig.7 confirms that the same
trend, and the same functional dependence for $\delta$G$_{\text{MD}}$/G(0) vs.
R$_{\square}$, is observed in GeSb$_{\text{x}}$Te$_{\text{y}}$ films.
\begin{figure}[ptb]%
\centering
\includegraphics[
height=2.3929in,
width=3.3399in
]%
{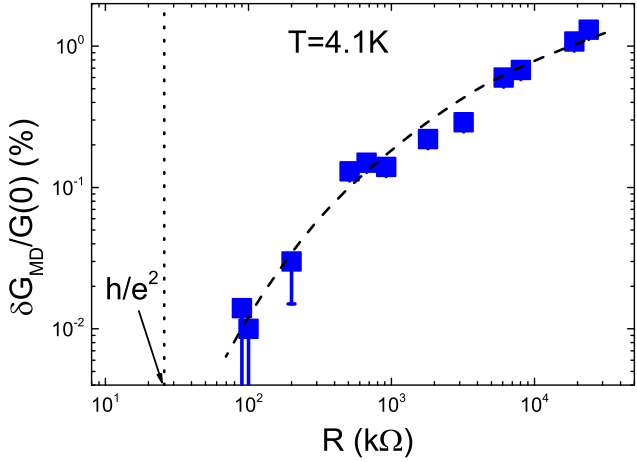}%
\caption{The dependence of the memory-dip magnitude $\delta$G$_{\text{MD}}%
/$G$_{\text{0}}$ on the film sheet-resistance R$_{\square}$. These were
obtained from G(V$_{\text{g}}$) curves of the different samples taken with the
same sweep rate dV$_{\text{g}}$/dt. The dashed curve through the data point is
guide to the eye. The dotted separates the diffusive regime from the strongly
localized one. A remarkably similar dependence on disorder was observed in
other electron-glasses \cite{34,35}. }%
\end{figure}

In contrast with the electron-glass phase that categorically requires
R$_{\square}$%
$>$%
h/e$^{\text{2}}$ for its existence, persistent-photoconductivity is observable
in the entire disorder regime covered in this work which includes samples in
the weakly-localized regime. The PPC relative magnitude $\delta$%
G$_{\text{PPC}}$/G$_{\text{0}}$ does depend on the film disorder as shown in
Fig.5 above but it is clearly observed in samples with R$_{\square}$%
$<$%
h/e$^{\text{2}}.$ In terms of the absolute value of the excess conductance it
is actually larger in diffusive samples than in the strongly-localized regime.
Once R$_{\square}$ becomes smaller than $\approx$10$^{\text{5}}\Omega$ the
magnitude of the MD quickly approaches the limit of the experimental
noise-level (Fig.7) while the excess-conductance associated with the PPC in
these GeSb$_{\text{x}}$Te$_{\text{y}}$ films is not affected by the impending
diffusive phase; neither the magnitude nor the parameters of the relaxation
law seem to show any qualitative change (Fig.5).

This observation can be illustrated by using a sample in this intermediate
disorder regime. Such data are shown in Fig.8.%
\begin{figure}[ptb]%
\centering
\includegraphics[
height=2.655in,
width=3.3399in
]%
{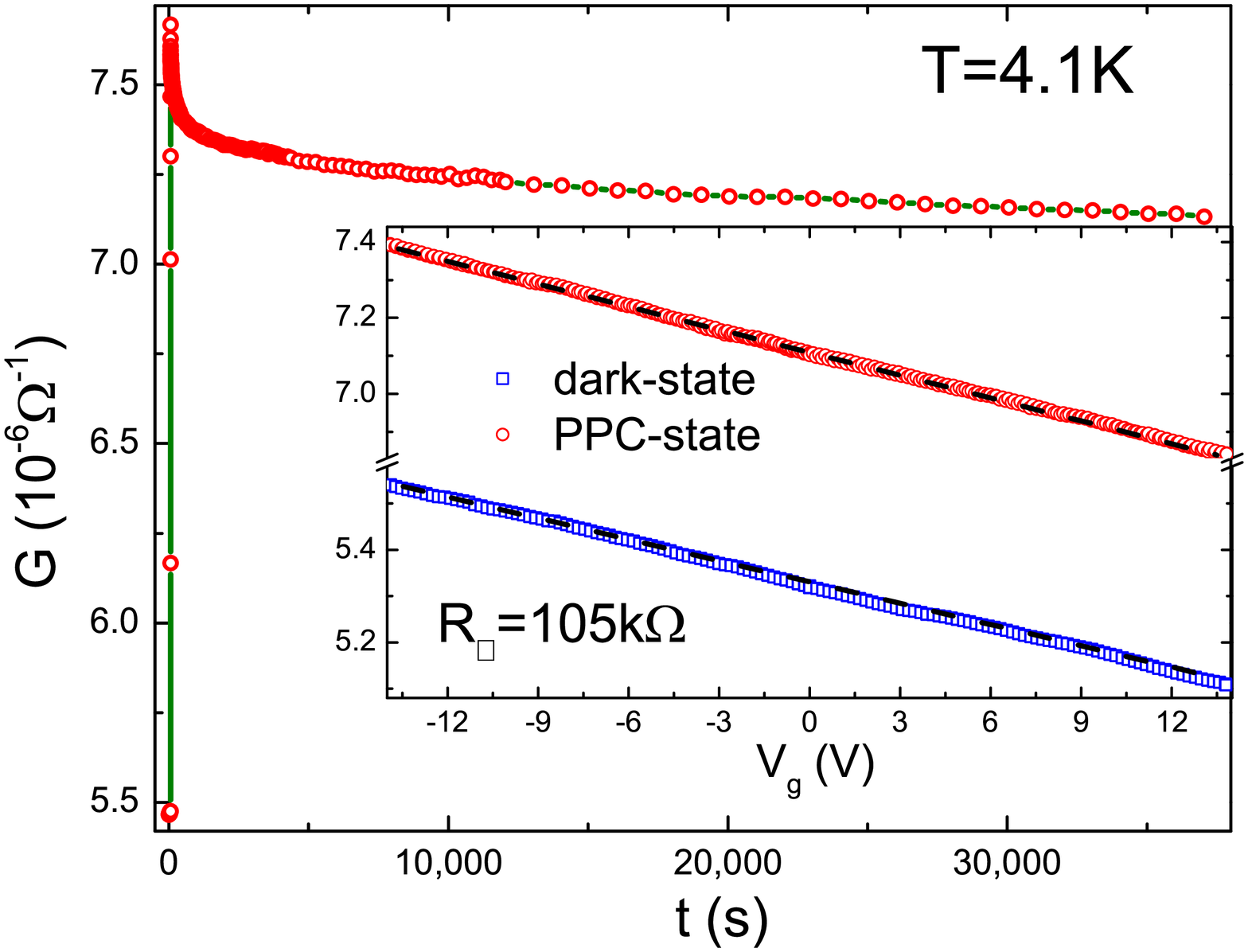}%
\caption{Comparing the result of the IR-protocol (main figure) with the
field-effect (inset) of a sample with an intermediate degree of disorder
(R$_{\square}$=105k$\Omega)$ before and after IR-illumination. The relaxation
of the conductance after the\ IR illumination is turned off is fitted by the
same stretched exponential dependence as the 19M$\Omega$ sample in Fig.3 with
$\beta$=0.1 and $\tau$=1$\cdot$10$^{\text{9}}$sec. Note that the field-effect
shows mainly a straight line with no clear signature of a memory-dip at
V$_{\text{g}}$=0V where the sample was equilibrated.}%
\end{figure}
The data in this figure demonstrate that an intermediate disorder regime may
be reached in a sample that still exhibits prominent PPC but the signature for
electron-glass is not to be seen above the noise level.

Obviously, strong-localization, a mandatory requirement for showing
electron-glass effects, is not a condition for observing
persistent-photoconductivity in GeSb$_{\text{x}}$Te$_{\text{y}}$. Empirically,
these two phenomena tend to favor vastly different carrier-concentration for
their observation; intrinsic electron-glass behavior, with relaxation times
larger than seconds, has been observed \textit{only} in systems with high
carrier-concentrations \cite{26}. Persistent-photoconductivity on the other
hand, has been reported almost exclusively in lightly-doped semiconductors.
Observation of the latter phenomenon requires in addition, the existence of a
specific defect-type to form a barrier that hinders fast recombination
\cite{8,13}.

These different preferences however, are apparently not mutually exclusive. As
shown next, the two phenomena may coexist.

\subsection{Lighting-up the Memory-Dip}

The existence of a memory-dip in GeSb$_{\text{x}}$Te$_{\text{y}}$ with
R$_{\square}$%
$>$%
10$^{\text{5}}\Omega$ was anticipated; these samples are Anderson-localized,
and they have relatively high carrier-concentration. These are the empirically
established conditions for observing the electron-glass phenomenon over
conveniently long time-scales \cite{26}. At this time there is no known
exception to this observation.

Exposing these samples to infrared radiation however, had an unexpected
effect. Figure 9 compares the field-effect measured on the sample that was
used in Fig.6 in the dark-state with a G(V$_{\text{g}}$) taken with the same
experimental protocol but after a brief IR exposure to set the system in its
PPC state. The latter G(V$_{\text{g}}$) reveals a \textit{more prominent} MD
than observed in the dark-state.%
\begin{figure}[ptb]%
\centering
\includegraphics[
height=2.5598in,
width=3.3399in
]%
{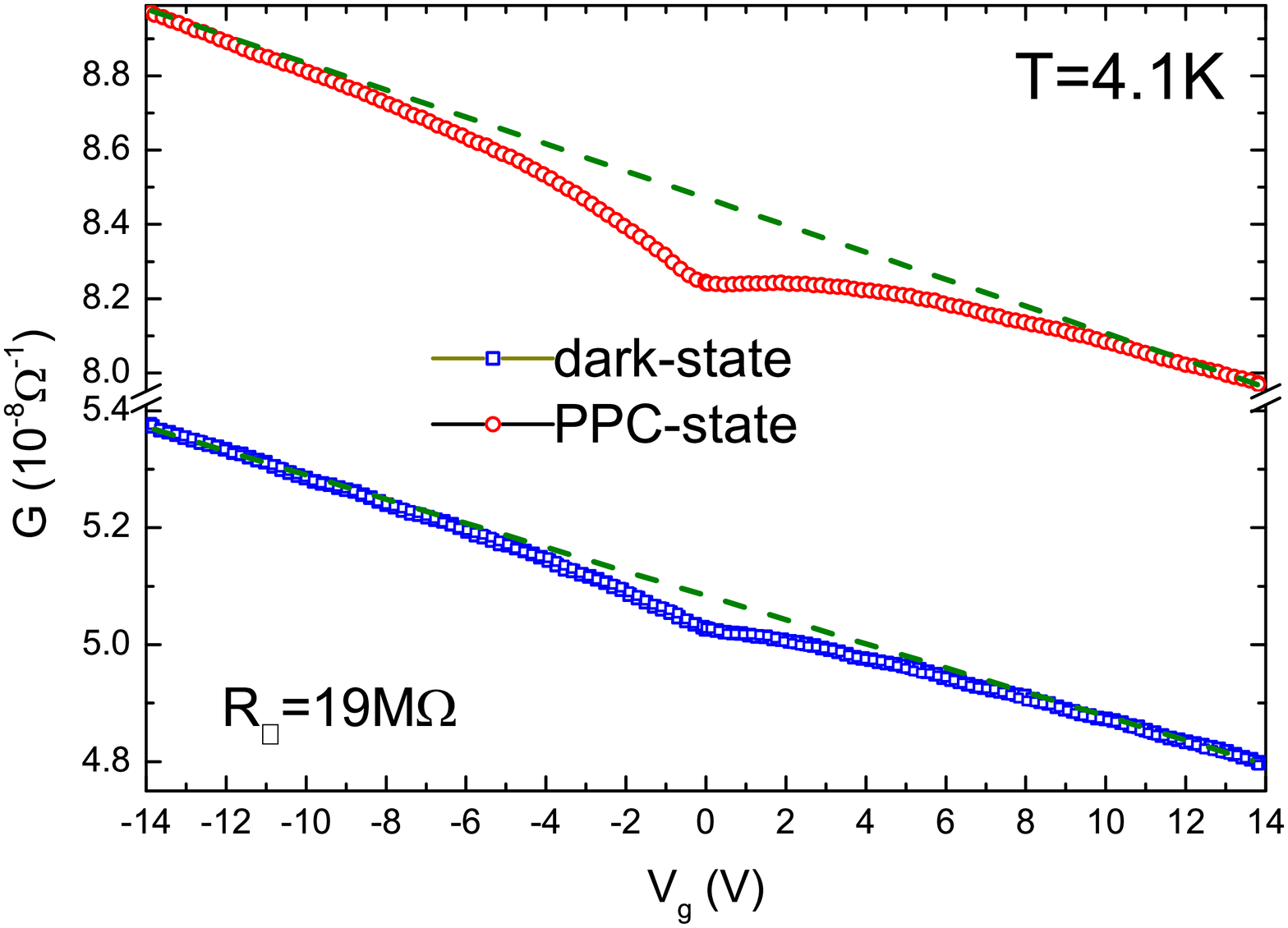}%
\caption{Comparing the field-effect characteristics G$_{\text{DS}}%
$(V$_{\text{g}}$) for the R$_{\square}$=19M$\Omega$ sample in the dark-state
with G$_{\text{IR}}$(V$_{\text{g}}$) taken one hour after exposing the sample
to 1mWatt IR radiation for 3sec (a $\approx$1 hour delay is necessary for
obtaining meaningful G(V$_{\text{g}}$) data; immediately after terminating the
illumination G is still varying significantly over the $\approx$1 minute
required for obtaining a useful G(V$_{\text{g}}$) reading across this range of
V$_{\text{g}}$). The figure clearly demonstrates that a more developed
memory-dip is developed in the irradiated sample. Dashed lines are the
thermodynamic component\ G(V$_{\text{g}}$) of the field-effect.}%
\end{figure}

Note that the conductance in this case increased by 60\% while the magnitude
of the memory-dip increased by 150\%, the absolute value of the MD component
$\delta$G$_{\text{MD}}$ increases by more than the average G. This
photo-enhanced memory-dip has been observed in \textit{all} our
GeSb$_{\text{x}}$Te$_{\text{y}}$ samples with R$_{\square}$%
$>$%
10$^{\text{5}}\Omega,$ and independent of the equilibrium value of
V$_{\text{g}}$in the dark-state within the range of -8V to +8V. It was never
encountered in previously studied electron-glasses where, if anything,
IR-exposure caused a \textit{diminishment} of the MD magnitude \cite{36,40}.
Actually in every instance where G increased, be it due to temperature,
non-Ohmic field, or decrease of disorder $\delta$G$_{\text{MD}}$/G
\textit{always wen}t \textit{down}.

Another deviation from the electron-glass norm is observed in the asymptotic
dependence of the MD amplitude on time. In previously studied electron-glasses
the MD amplitude was found to \textit{grow} logarithmically with the time the
system relaxes under a fixed V$_{\text{g}}$ \cite{35}. In the PPC state
however, the MD amplitude actually \textit{decreases} at long times. The
example shown in Fig.10 illustrates how, concomitant with the slow decay of
the PPC, the MD magnitude diminishes with time.%
\begin{figure}[ptb]%
\centering
\includegraphics[
height=2.8513in,
width=3.3399in
]%
{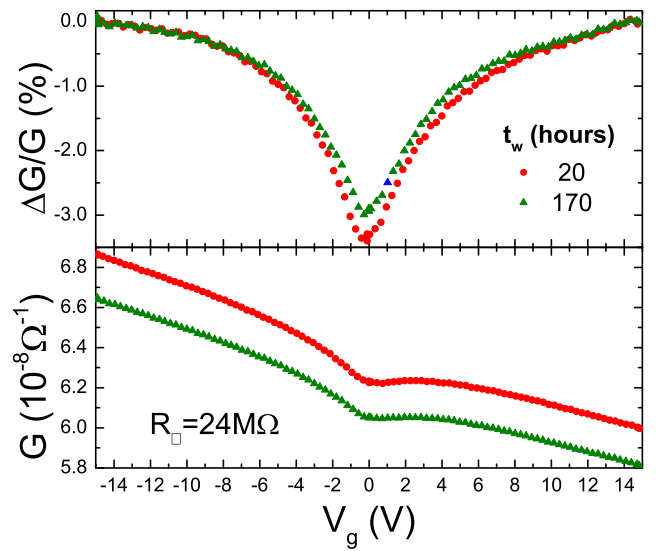}%
\caption{The asymptotic behavior of the field-effect versus the time elapsed
since put in the PPC-state. Bottom graph shows the full G(V$_{\text{g}}$)
curves. Upper graph illustrates that the magnitude of the memory-dip
diminishes at long times. The curves in the upper graph were obtained form the
G(V$_{\text{g}}$) data \ in the bottom graph by subtracting the linear part of
the field-effect).}%
\end{figure}

These observations lead us to consider the following scenario to account for
this seemingly paradoxical IR effect.

Let us assume that the sample is not uniform; some regions in it are not part
of the electron-glass phase. These could be of two types; regions where the
disorder is weak such that locally they are diffusive, and regions in the
sample where the carrier-concentration is low. Neither type of a region would
exhibit long-lasting electron-glass dynamics if probed by itself; conspicuous
electron-glass effects in field-effect experiments are only detectable in
systems where both, high carrier-concentration \textit{and} strong disorder
are present \cite{26}. Therefore, regions that do not meet this requirement
would exhibit just a linear G(V$_{\text{g}}$), reflecting the material
thermodynamic DOS dependence on energy, but no memory-dip. On the other hand,
they are just as good as other regions to contribute to PPC. Actually, all
other things being equal, regions with low carrier-concentration are more
likely to be significantly affected by the optical excitation. Once primed by
the IR, some of these regions will just exhibit smaller resistance. Some
others, having sufficiently large resistance may have their
carrier-concentration elevated enough to exhibit memory-dip thus adding weight
to the global component. Either way, a larger magnitude of the MD should
result in the PPC-state.%
\begin{figure}[ptb]%
\centering
\includegraphics[
height=2.8513in,
width=3.3399in
]%
{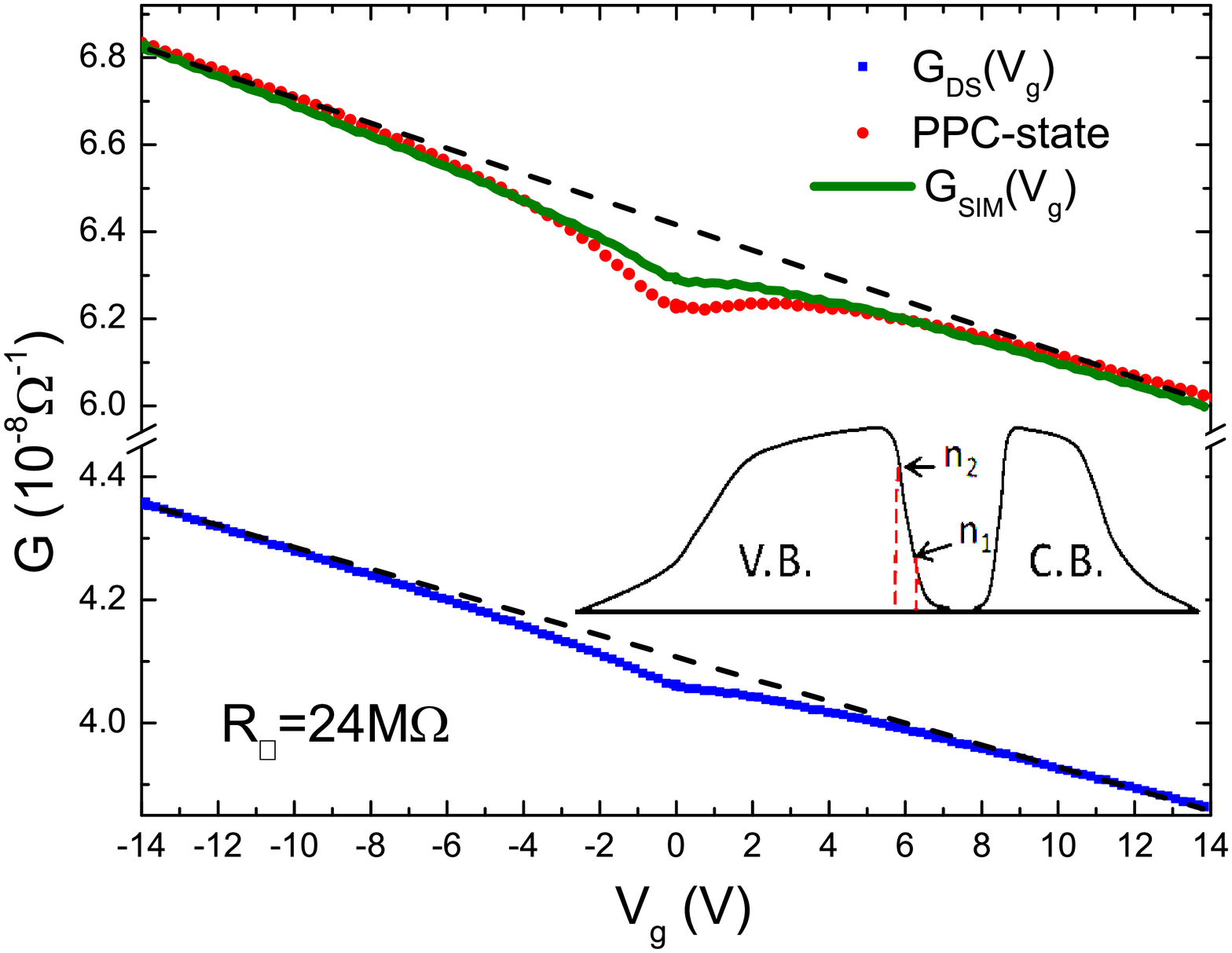}%
\caption{The field-effect for the 24M$\Omega$ sample in its dark-state
(squares) and in the PPC-state (circles). Dashed lines are the thermodynamic
component of G(V$_{\text{g}}$). The full curve labeled G$_{\text{SIM}}%
$(V$_{\text{g}}$), is the simulated curve (see text for details). This
particular sample was chosen for this test due to its large MD amplitude in
the dark-state but the same qualitative outcome was observed with other
samples. The schematic overlay depicts a "smoothed" density of states for the
valence V.B. and conduction band V.B. respectively (following the
band-structure calculation of reference \cite{38}). This illustrates why the
thermodynamic component part of the field-effect is sloped as in our
experiments for both, the state with carrier-concentration n$_{\text{1}}$ as
in n$_{\text{2}}$. }%
\end{figure}

A simple check on whether this scenario might be in the right direction to
account for the IR-enhancement effect is illustrated in Fig.11. The resistance
of the alleged non-glass regions is taken as in series with the electron-glass
part of the sample. The field-effect that results in the PPC state,
G$_{\text{SIM}}$(V$_{\text{g}}$) is then estimated by: G$_{\text{SIM}%
}^{\text{-1}}$(V$_{\text{g}}$)=G$_{\text{DS}}^{\text{-1}}$(V$_{\text{g}}%
$)-G$_{\text{PAR}}^{\text{-1}}$(V$_{\text{g}}$) where G$_{\text{DS}}%
$(V$_{\text{g}}$) is the experimentally measured field-effect in the
dark-state, and $\delta$G$_{\text{PAR}}^{\text{-1}}$(V$_{\text{g}}$) is the
resistance change of the non-glass regions when their dark
carrier-concentration n$_{\text{1}}$ is increased to n$_{\text{2}}$ by the IR
exposure (see schematic description in Fig.11). The model uses 2 parameters;
the slope and the value of G$_{\text{PAR}}$(V$_{\text{g}}$) at V$_{\text{g}}%
$=0. These are varied to match the overall slope of the experimental
G(V$_{\text{g}}$) in the PPC state. This simple modeling already accounts for
a large part of the MD-enhanced effect. To account for the full enhancement it
may be necessary to include the contribution from regions that are turned into
the electron-glass state when primed by the IR, which will also involve a more
refined treatment, probably invoking a percolation model of the inhomogeneous
scenario than the simple procedure used above.

There is hardly a need to justify the assumption of inhomogeneity of these
samples, being disordered is tantamount to being inhomogeneous. The question
is how the inhomogeneity manifests itself; what physical property varies in
space and above what spatial scale can the system be considered homogeneous?
In the present case it is of interest to assess the contribution of
fluctuations in carrier-concentration to the quenched disorder.

Spatial variations of carrier-concentration is not uncommon in multi-component
systems. It has been recognized for example, as the reason for
magneto-transport anomalies in Hg$_{\text{1-x}}$Cd$_{\text{x}}$Te films
\cite{41} (also a chalcogenide compound). Palm \textit{et al} referred to the
nature of inhomogeneities in their Hg$_{\text{1-x}}$Cd$_{\text{x}}$Te samples
by: "...different amounts of the insulator's fixed charge in various
regions..." stating that ..."this is like\ having different devices within one
sample..." \cite{41}.

This type of inhomogeneity appears also in amorphous compounds such as
indium-oxide. As pointed out by Givan and Ovadyahu \cite{42}, the energy cost
of forming charge-gradients in a conducting matter may be mitigated by
variation in chemical composition. Inhomogeneous carrier-concentration is
especially important when a many-body phenomenon like superconductivity is
involved \cite{42}. The sensitivity to carrier-concentration \cite{26,38}
makes it also important for the electron-glass phenomenon.

GeTe-based compounds are as susceptible to this kind of disorder as any other
multi-component system. Using heat-treatment, Bahl and Chopra were able to
vary the carrier-concentration in their GeTe samples by more than an order of
magnitude \cite{43} while preserving the same crystalline symmetry of the
material. Crystal chemistry constraints in these materials are therefore not
very effective in suppressing stoichiometry fluctuations. This should be
compared with the mere factor of 2 variation in carrier-concentration possible
in (self-doped) crystalline indium-oxide In$_{\text{2}}$O$_{\text{3-x}}$
without doping by foreign elements \cite{40}. Adding a third element (Sb) to
the GeTe system apparently only allows for greater variability; reported
values for n in GeSb$_{\text{x}}$Te$_{\text{y}}$ with various x and y range
between $\approx$3$\cdot$10$^{\text{18}}$cm$^{\text{-3}}$ \cite{32} to
$\approx$10$^{\text{22}}$cm$^{\text{-3}}$\cite{29}. Also, the large dielectric
constant of GeSb$_{\text{x}}$Te$_{\text{y}}$ \cite{28} further reduces the
energy-cost associated with carrier-concentrations gradients.

To examine the spatial extent of these inhomogeneities, we initiated a study
of the variation in the local-stoichiometry in GeSb$_{\text{x}}$Te$_{\text{y}%
}$ films. Recall that these materials are conducting by virtue of vacancies
\cite{39} and therefore we expect that mapping stoichiometry may be indicative
of local carrier-concentration. This was done by
employing\ energy-dispersive-spectroscopy (EDS) in the STEM mode at 200kV
beam. The methodology is based on building the distribution of the local
stoichiometry by performing consecutive EDS reading over a square with lateral
side L. The study revealed $\approx$20\% variations in both Ge:Te and Sb:Te
atomic-ratios in a distribution based on stoichiometry readings of squares
with L=400nm taken across a given sample. A spatial scale of 400nm is much
larger than typical values of the hopping-length in strongly-localized samples
at $\simeq$4K \cite{44}. This makes the inhomogeneous scenario plausible; the
spatial scale associated with the compositional disorder is large enough to
justify treating the sample as being composed of segments with different
transport characteristics.

It is natural to ask why this IR-enhancement effect has not been observed, for
example, in amorphous indium-oxide where compositional disorder similar to
that of GeSb$_{\text{x}}$Te$_{\text{y}}$ has been found \cite{42}. Actually,
at least to some degree the effect might be expected to exist in \textit{any}
electron-glass: These systems are \textit{inherently} inhomogeneous and the
notion of regions in the current-carrying network being in series with
bottleneck resistances is a natural part of percolation models
\cite{45,46,47,48} that purport to describe their transport properties.

A likely reason for the absence of this effect in the previously studied
electron-glasses is simple: they do not exhibit PPC. To show
persistent-photoconductivity there are rather restrictive conditions to be
met. The first is feasibility of charge-generation from the light-source. The
optical gap of GeSb$_{\text{x}}$Te$_{\text{y}}$ is 0.4-0.8~eV \cite{27}
allowing \textit{inter}-band transitions which evidently involve generation of
excess charge in the system (note however that effective PPC may be achieved
sometimes even for smaller photon energies than necessary for interband
transitions \cite{49}). Indium-oxide and other previously studied
electron-glasses, tested with the same IR-protocol as in this work, do not
meet this requirement. The photon energy of the IR source used in these
experiments ($\approx$1eV) is smaller than their optical gap \cite{30,50}.
This allows only \textit{intra}-band transitions, which may be effective in
randomizing site-occupation which (under sufficiently large energy flux),
diminish the MD \cite{36,40}. However, no extra charge is generated in the
system under these conditions.

A long-lasting PPC state is not guaranteed even when using a source with a
photon energy that exceeds the optical gap. To keep the photo-generated
electron-hole pairs from recombining and preserve the excess conductance, a
mechanism has to exist to create an effective barrier against fast
recombination. In general it is hard to identify this mechanism in a given
system \cite{7}. It has not yet been identified in GeSb$_{\text{x}}%
$Te$_{\text{y}}$ either. It would be interesting to extend the study of
infrared excitation to GeTe samples, eliminating the Sb, which may help in
identifying the defect responsible for in the persistent-photoconductivity we
observe in the GeSbTe compounds.

As already mentioned, PPC has been observed in many systems but, to our
knowledge, not in a system with carrier-concentration as high as in these
GeSb$_{\text{x}}$Te$_{\text{y}}$ films. Enhancing the conductance of a system
with n$\approx$10$^{\text{20}}$-10$^{\text{21}}$cm$^{\text{-3}}$ by 10-100\%
(Fig.6) seems to require a photoinduced $\delta$n of not smaller smaller than
$\approx$10$^{\text{19}}$-10$^{\text{20}}$cm$^{\text{-3}}.$ This may be
estimated by comparison to field-effect data; $\delta$G(V$_{\text{g}}$)
associated with the thermodynamic component results from adding (removing) a
charge of $\delta$n to (from) the sample. This charge is determined by
$\delta$V$_{\text{g}}$ and the sample-gate capacitance C. For $\delta
$V$_{\text{g}}$=20V, and C$\approx$10nF$\cdot$cm$^{\text{-2}},$ the extra
charge $\delta$n typically associated with the range of our G(V$_{\text{g}}$)
measurements is of order 10$^{\text{18}}$-10$^{\text{19}}$cm$^{\text{-3}}$
(depending on whether the charge is confined to a screening-layer or is spread
evenly along the film thickness). The conductance change $\delta$G associated
with this $\delta$n varies between 0.7\% to 9.5\% for samples with
R$_{\square}$=2.5k$\Omega$ and R$_{\square}$=24M$\Omega$ respectively. The
conductance increment in the PPC state $\delta$G$_{\text{IR}}$ is an order of
magnitude larger (see Fig.6), which may suggest that the associated $\delta$n
is also larger. Perhaps much larger.

To put things is perspective, it is interesting to compare our PPC results
with those obtained for another chalcogenide that is well known to exhibit
persistent-photoconductivity, Pb$_{\text{1-x}}$Sn$_{\text{x}}$Te. This
compound has similar material characteristics as GeSb$_{\text{x}}%
$Te$_{\text{y}}$ (chemistry, optical-gap, large dielectric-constant, p-type
conductivity). Persistent-photoconductivity in Pb$_{\text{1-x}}$Sn$_{\text{x}%
}$Te compounds doped with In or Ga has been extensively studied \cite{7}.
Working with bulk crystals allowed the researchers an effective way to
estimate the photoinduced excess charge $\delta$n by analyzing Shubnikov-de
Haas oscillations data. Values for $\delta$n, under energy-flux levels of
similar magnitude as used in this work, reached only 10$^{\text{17}}%
$-10$^{\text{18}}$cm$^{\text{-3}}$ \cite{7}. Such values appear to be too low
to account for our PPC results. This may indicate that photo-generation of
charge is more efficient in GeSb$_{\text{x}}$Te$_{\text{y}}$ or that the two
processes of inserting the same $\delta$n to the system affect its conductance
differently, perhaps due to the inhomogeneity (an inherent property of the
hopping systems). These issues must await a direct measurement of the
photoinduced excess-charge in GeSb$_{\text{x}}$Te$_{\text{y}}$ films.

In sum, we presented in this paper experimental results that demonstrate the
existence of persistent-photoconductivity in the degenerate semiconductor
GeSb$_{\text{x}}$Te$_{\text{y}}$. This non-stoichiometric compound has rather
high carrier-concentration and thus it also exhibits electron-glass effects
when in the strongly localized regime. It is demonstrated that
persistent-photoconductivity and electron-glass phase are different
nonequilibrium phenomena. They manifestly may coexist\ in the strongly
localized regime of GeSb$_{\text{x}}$Te$_{\text{y}}$ although it is not
entirely clear that they share the same spatial regions in the disordered
system. The memory-dip, a distinguishing feature of the electron-glass, is
appreciably enhanced In the PPC-state. There is also an intriguing interplay
between these phenomena in terms of a dramatic slowdown of the electron-glass
dynamics while in the PPC-state that will be treated in future work.

Further work is needed to study the detailed temperature dependence of the
persistent-photoconductivity in these materials and to elucidate the nature of
the defects that are associated with this effect.

\begin{acknowledgments}
Illuminating discussions with Dmitry Khokhlov on persistent-photoconductivity
are gratefully acknowledged. This research has been supported by a grant
administered by the Israel Academy for Sciences and Humanities.
\end{acknowledgments}


\begin{thebibliography}{99}
\bibitem {44}Typical values for hopping length at T$\approx$4K for a variable
range hopping system is rarely larger than $\approx$50nm.
\end{thebibliography}
\end{document}